\documentclass[a4paper,fleqn]{cas-dc}

\usepackage[numbers]{natbib}
\usepackage{graphicx}%
\usepackage{multirow}%
\usepackage{amsmath,amssymb,amsfonts}%
\usepackage{mathtools}
\usepackage{amsthm}%
\usepackage{mathrsfs}%
\usepackage[title]{appendix}%
\usepackage{xcolor}%
\usepackage{textcomp}%
\usepackage{manyfoot}%
\usepackage{booktabs}%
\usepackage{algorithm}%
\usepackage{algorithmicx}%
\usepackage{algpseudocode}%
\usepackage{listings}%
\usepackage{widetext}
\usepackage{ulem}
\usepackage{cancel}

\definecolor{magenta}{rgb}{0.8, 0.1, 0.6}

\def\v#1{{\boldsymbol{#1}}} 
\def\m#1{{\mathcal{#1}}} 

\def\tsc#1{\csdef{#1}{\textsc{\lowercase{#1}}\xspace}}
\tsc{WGM}
\tsc{QE}
\tsc{EP}
\tsc{PMS}
\tsc{BEC}
\tsc{DE}

\begin{document}
\let\WriteBookmarks\relax
\def\floatpagepagefraction{1}
\def\textpagefraction{.001}
\shorttitle{Viscoelasticity of two-dimensional fluid membrane}
\shortauthors{N. Takeishi \& N. Yokoyama et~al.}

\title[mode = title]{Numerical analysis of viscoelasticity of two-dimensional fluid membranes under oscillatory loadings}


\author[1,2]{Naoki Takeishi}[
   type=editor,
   auid=000, bioid=1,
   orcid=0000-0002-9568-8711
   ]
\cormark[1]
\ead{ntakeishi@kit.ac.jp}

\author[2]{Masaya Santo}[
   ]

\author[3]{Naoto Yokoyama}[
   orcid=0000-0003-1460-1002
   ]
\cormark[1]
\ead{n.yokoyama@mail.dendai.ac.jp}

\author[2]{Shigeo Wada}


\address[1]{Department of Mechanical Engineering, Kyoto Institute of Technology, Goshokaido-cho, Matsugasaki, Sakyo-ku, 606-8585, Kyoto, Japan}
\address[2]{Graduate School of Engineering Science, Osaka University, 1-3 Machikaneyama, Toyonaka, 560-8531, Osaka, Japan}
\address[3]{Department of Mechanical Engineering, Tokyo Denki University, 5 Senju-Asahi, Adachi, 120-8551, Tokyo, Japan}

\cortext[cor1]{Corresponding author}


\begin{abstract}
Biomembranes consisting of two opposing phospholipid monolayers, which comprise the so-called lipid bilayer, are largely responsible for the dual solid-fluid behavior of individual cells and viruses.
Quantifying the mechanical characteristics of biomembrane, including the dynamics of their in-plane fluidity, can provide insight not only into active or passive cell behaviors but also into vesicle design for drug delivery systems.
Despite numerous studies on the mechanics of biomembranes,
their dynamical viscoelastic properties have not yet been fully described.
We thus quantify their viscoelasticity based on a two-dimensional ($2$D) fluid membrane model,
and investigate this viscoelasticity under small amplitude oscillatory loadings in micron-scale membrane area.
We use hydrodynamic equations of bilayer membranes,
obtained by Onsager's variational principle,
wherein the fluid membrane is assumed to be an almost planar bilayer membrane.
Simulations are performed for a wide range of oscillatory frequencies $f$ and membrane tensions.
Our numerical results show that as frequencies increase,
membrane characteristics shift from an elastic-dominant to viscous-dominant state.
Furthermore, such state transitions obtained with a $1$-$\mu$m-wide loading profile appear with frequencies between $O(f) = 10^1$--$10^2$ Hz, 
and almost independently of surface tensions.
We discuss the formation mechanism of the viscous- or elastic-dominant transition based on relaxation rates that correspond to the eigenvalues of the dynamical matrix in the governing equations.
\end{abstract}




\maketitle

\section{Introduction}\label{sec1}

Lipid bilayer membranes,
consisting of a series of opposing phospholipids arranged in a two-dimensional ($2$D) fluid crystalline assembly with $\sim 5$-nm thickness~\citep{Singer1972},
are a common and fundamentally important structure in all living cells and many viruses~\citep{Lamers2022}.
Each lipid covers a surface area of approximately $0.7$ nm$^2$ ($= 70$ {\AA}$^2$)~\citep{Engelman1969}.
The membrane structure separates the inside and outside of the cell,
and assumes various function-related shapes~\citep{Zimmerberg2006}.
In addition,
membrane mechanical properties affect cell and membrane dynamics,
such as active cell migration~\citep{Janmey2006} and endocytosis~\citep{Anishkin2014, DizMunoz2013}. 
From a mechanical viewpoint,
while phospholipids in membranes can move in the planar direction,
their displacement in their vertical displacement is restricted,
and thus the bilayers can behave as a $2$D fluid membrane.
Such fluid deformable surfaces exhibit a solid-fluid duality,
resulting in unique and complex mechanical characteristics wherein in-plane fluidity and elasticity can emerge simultaneously.
Thus, quantifying the dynamical mechanical properties of biomembranes can provide insight not only into the aforementioned cell behaviors but also into the development of drug delivery systems with vesicles~\citep{Herrmann2021},
which are closed biomembranes that typically consist of only lipids and cholesterol without any proteins.

Although the mechanics of biomembranes have been well studied using various experimental strategies such as atomic force microscopy (AFM),
micropipette aspiration,
and optical tweezers,
as reviewed in Ref.~\citep{Suresh2007},
the dynamical viscoelasticity of (pure) lipid bilayers under oscillatory loadings has not yet been fully understood.
Recent experimental techniques have successfully quantified dynamical viscoelasticity by complex moduli in lipid monolayers or bilayers,
assuming a linear mechanical response of membranes to oscillatory shear strains~\citep{Al-Rekabi2018, Choi2011, Harland2011, Kim2011}.
For instance,
for different concentrations of cholesterol [i.e., mixtures of $1$,$2$-dipalmitoyl-$sn$-glycero-$3$-phosphocholine (DPPC) and cholesterol],
Al-Rekabi et al. used AFM to produce a map of the viscoelastic properties of a lipid bilayer composed of DPPC\citep{Al-Rekabi2018},
which is one of the primary lipids in lung surfactant~\citep{Lipp1996} and is ubiquitous in cell membranes.
Despite these efforts, there is still no consensus on the transition mechanism between viscous- and elastic-dominant states in lipid bilayers.

Along with these experimental studies,
various theoretical frameworks have been proposed to describe fluid membrane dynamics~\citep{Barthes-Biesel1985, Secomb1982, Seifert1997},
and some have been applied to problems regarding the spontaneous conformation of human red blood cells (RBCs)~\citep{Lim2002} and vesicles~\citep{Dobereiner2000}.
In these works,
the lipid bilayer is modeled as a continuous elastic surface~\citep{Helfrich1973},
considering the scale difference between the micrometer system size and nanometer membrane thickness.
Lipowsky described spontaneous curvature by an approximation of solid shells that store elastic energy during stretching or bending~\citep{Lipowsky1991}.
In terms of soft matter physics,
Seifert and Langer successfully described bilayer hydrodynamics for almost planar membranes~\citep{Seifert1993},
where coupling of the membrane dynamics with the surrounding fluid was taken into account by modeling curvature,
density-difference elasticity,
intermonolayer friction,
monolayer $2$D viscosity,
and solvent three-dimensional ($3$D) viscosity.

To further provide a precise mechanical background in membrane dynamics,
some numerical models based on the aforementioned results of \citep{Seifert1993} have been proposed,
which make it possible to deal with the complex interplay between membrane elasticity and hydrodynamic forces acting at microscopic scales.
Fournier use Onsager's variational principle to derive the governing equations describing the dynamics of an almost planar bilayer membrane~\citep{Fournier2015},
and numerically investigated the effect of membrane tension on the relaxation rate.
The principle here is an established, unified framework for the dissipative dynamics of a soft matter system~\citep{Doi2011, Goldstein2001}.
It provides hydrodynamical equations pertaining to bilayer membranes by minimizing a Rayleighian consisting of potential power for dynamical changes and of dissipated power,
to resist the change.
More recently,
Torres et al. proposed new computational methods that build on Onsager's formalism and arbitrarily Lagrangian-Eulerian formulations~\citep{Torres2019}.
Their methodologies were successfully applied not only to dynamic lipid bilayers,
but also to adhesion-independent cell migration.
A molecular dynamic (MD) approach has been applied to the oscillatory behavior of the lipid bilayer membrane~\citep{Li2020} and to membrane fluctuations of RBCs~\citep{Sadeghi2020}.
It is expected that the membrane viscoeasticity would actually be observed in all-atom MD simulations as an emergent property if viscoelastic effect are innate to the bilayer.
However, this has not yet achieved due to massive computational load.
Thus, despite these efforts,
the dynamical viscoelastic nature of lipid bilayers,
especially with regard to tensile loadings,
has not yet been fully described.

Therefore, the objective of this study is to clarify whether the lipid bilayer characterized by a $2$D fluid membrane featuring lipid bilayers exhibits a transition between the viscous- and elastic-dominant states depending on the oscillatory loading frequencies.
We have made use of the AFM measurements in DPPC bilayers~\citep{Al-Rekabi2018, Rigato2017} in our simulations.
In order to limit the study to linear mechanical response,
small amplitude oscillatory tensile loadings are considered.
If frequency-dependent viscoelasticity exhibits in the $2$D fluid membrane,
we also discuss whether the formation mechanism of the viscous- or elastic-dominant transition can be explained based on relaxation rates that correspond to the eigenvalues of the dynamical matrix in the governing equations.
The theoretical framework of $2$D fluid membrane in the present study follows the previous study by \cite{Fournier2015}.
To quantify the dynamical viscoelasticity of the $2$D fluid membrane model,
we propose metrics evaluated by scaled mass density and stress in the membrane:
the complex moduli $E^\ast (\omega) = E^\prime (\omega) + i E^{\prime\prime} (\omega)$,
where $i = \sqrt{-1}$ is the imaginary unit,
$E^\prime$ is the storage modulus representing the elastic component of the stress,
and $E^{\prime\prime}$ is representing the viscous dissipation.
In this study,
in terms of the way of quantification,
we distinguish $E^\ast$ and classical viscoelastic metrics as complex ``shear'' moduli,
$G^\ast (\omega) = G^\prime (\omega) + i G^{\prime\prime} (\omega)$ although these may be potentially equivalent.
Simulations are performed for wide range of loading frequencies $\omega$ ($= 2 \pi f$) and surface tensions $\sigma$.

\section{Methods}\label{sec2}
\subsection{Model of lipid bilayer membrane}\label{subsec2_1}

\begin{figure}[htbp]
  \centering
    \includegraphics[clip,width=7cm]{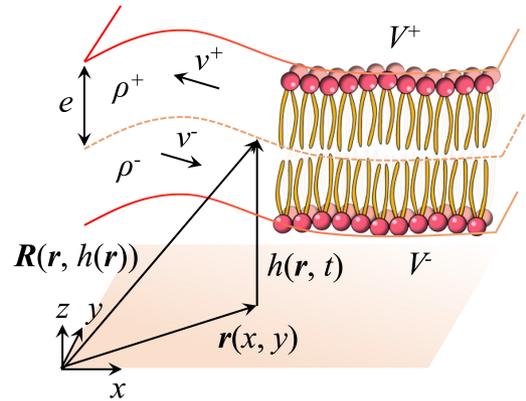}
    \caption{
	    Geometrical description of the membrane shape $h (\v{r}, t)$ and membrane coordinate $\v{R} (\v{r}, h(\v{r}))$ with scaled densities $\rho^{\pm}$,
	    monolayer velocities $v^{\pm}$,
	    bulk solvent velocities $V^{\pm}$,
    	    and the surface distance away from the membrane mid-surface $e$.
    }
    \label{fig:model}
\end{figure}

Following the previous theoretical and numerical study by Fournier~\citep{Fournier2015},
we consider a lipid bilayer membrane made of only one lipid type in an unbounded flow field.
The membrane shape is therefore characterized by the height $z = h (\v{r}, t)$ from the plane at $z = 0$ to the membrane mid-surface,
where $\v{r}$ is the membrane coordinate projected onto the $x$-$y$ plane.
Position of a point on the membrane in this coordinate system is thus denoted as $\v{R} = (\v{r}, h(\v{r}))$.
Two monolayers in the membrane possess a lipid mass density $n^\pm (\v{r}, t)$ in deformed state $h (\v{r}, t)$,
where superscript ``$\pm$'' represents the upper monolayer ($z > h$) and lower monolayer ($z < h$), respectively.
Using mass density $n_0$ in the tensionless state as the reference, the membrane state can be described by its shape (or height) $h (\v{r}, t)$ and scaled mass density $\rho^{\pm} (\v{r}, t)$ as:
\begin{equation}
	\rho^{\pm} (\v{r}, t)
	= \frac{n^{\pm} (\v{r}, t) - n_0}{n_0}.
  \label{rho}
\end{equation}
We also consider the $3$D solvent velocities $V_\alpha^\pm (\v{R}, t)$ ($\alpha = x$, $y$, $z$) on either side of the membrane,
and the $2$D lipid velocities $v_i^\pm (\v{r}, t)$, ($i = x$,  $y$) in both monolayers.
A schematic of the $2$D fluid membrane is shown in Fig.~\ref{fig:model}.

The bulk solvent is assumed to be an incompressible fluid,
and hence $V_\alpha^\pm$ satisfies the following mass conservation law:
\begin{equation}
	\partial_\alpha V_\alpha^{\pm} = 0,
  \label{mass_flu}
\end{equation}
where $\partial_\alpha = \partial/\partial R_\alpha$.
The lipid mass conservation equation is $\partial_t n^\pm + \partial_i (n^\pm v_i^\pm) = 0$, where $\partial_i = \partial/\partial r_i$.
The first order approximation leads to the following equation:
\begin{equation}
\partial_t \rho^{\pm} + \partial_i v_i^{\pm} = 0.
  \label{mass_mem}
\end{equation}
The Einstein summation convention is adopted,
wherein repeating the same index twice in a single term implies summation over all possible values of that index.
Because of no-slip boundary conditions on the membrane surface,
the external fluid must satisfy the following equations: 
\begin{align}
	&V_i^{\pm}|_{z = h} = v_i^{\pm}, \label{no-slip_xy} \\
	&V_z^{\pm}|_{z = h} = \partial_t h. \label{no-slip_z}
\end{align}
Equation~\eqref{no-slip_xy} gives
\begin{align}
 \partial_i v_i = \left. \partial_i V_i^{\pm} \right|_{z=h} + \left. \frac{\partial V_i}{\partial z} \right|_{z=h} \partial_i h
,
 \label{partial_v}
\end{align}
and the first term of the right-hand side is equal to $- \left. \left(\partial V_z^\pm/\partial z \right) \right|_{z = h}$ due to Eq.~\eqref{mass_flu}.
The first order approximation as well as the Stokes approximation allows us to neglect the second term in Eq.~\eqref{partial_v}.
Then, Eq.~\eqref{mass_mem} gives
\begin{align}
\rho^\pm = \int \left. \frac{\partial V_z^\pm}{\partial z} \right|_{z = h} dt
.
      \label{strain_zz}
\end{align}
Because $\left( \rho^+ + \rho^- \right)$ relaxes almost instantly to zero,
the upper and lower monolayers satisfy $\rho^+ + \rho^- \simeq 0$~\citep{Fournier2015}.
More precise derivations of Eqs.~\eqref{partial_v} and \eqref{strain_zz} are described in Appendix~\S\ref{secA1}.

\subsection{Rayleighian and power components}\label{subsec2_2}
In the Stokes approximation (i.e., all forces balance without all inertial effects),
the dynamical equations for the motion of the membrane in the bulk solvent can be given by minimizing the total Rayleighian of the system with respect to all the dynamical variables~\citep{Arroyo2009, Doi2011, Rahimi2012}:
\begin{align}
	\m{R}
	&= \frac{1}{2} W + \dot{\m{H}} \\
	&= \m{P}_b^\pm + \m{P}_s^\pm + \m{P}_i^\pm + \dot{\m{H}}_{int} + \dot{\m{H}}_{ext}.
       \label{R}
\end{align}
Here, $W/2 (= \m{P}_b^\pm + \m{P}_s^\pm + \m{P}_i^\pm)$ is the resistive power against dynamical changes,
the so-called dissipated power,
which consists of three sources~\citep{Goldstein2001}:
$\m{P}_b$, the viscous dissipation in the bulk solvent above and below the membrane;
$\m{P}_s$, the viscous dissipation in the lipid fluids of the two monolayers due to $2$D viscosity;
and $\m{P}_i$, the dissipation associated with the intermonolayer friction~\citep{Evans1994, Seifert1993}.
Moreover, $\dot{\m{H}}$ ($= \dot{\m{H}}_{int} + \dot{\m{H}}_{ext}$) is the driving power for dynamical changes and consists of the intrinsic elastic power $\dot{\m{H}}_{int}$ and the external elastic power $\dot{\m{H}}_{ext}$.

As described in Ref.~\citep{Fournier2015}, 
the three dissipated power sources in Eq.~\eqref{R} can be written as:
\begin{align}
	&\m{P}_b^\pm = \int_{B^\pm} d\v{R} \eta D_{\alpha\beta}^\pm D_{\alpha\beta}^\pm, \\
	&\m{P}_s^\pm = \int d \v{r} \left( \eta_2 d_{ij}^\pm d_{ij}^\pm + \frac{\lambda_2}{2} d_{ii}^\pm d_{jj}^\pm \right), \\
	&\m{P}_i^\pm = \int d \v{r} \frac{b}{2} \left( \v{v}^+ - \v{v}^- \right)^2,
  \label{e_source1}
\end{align}
where $B^\pm$ is the volume defined by $z > h$ or $z < h$,
$D_{\alpha\beta}^\pm (\v{r}) = \left( \partial_\alpha V_\beta^\pm + \partial_\beta V_\alpha^\pm \right)/2$ is the rate-of-deformation tensor in the bulk solvent,
$d_{ij}^{\pm} (\v{r}_s) = \left( \partial_i v_j^\pm + \partial_j v_i^\pm \right)/2$ is the rate-of-deformation tensor in the monolayer fluids,
$\eta$ is the bulk solvent viscosity,
$\eta_2$ is the $2$D viscosity,
$\lambda_2$ is the dilational viscosity,
and $b$ is the intermonolayer friction coefficient.

Assuming a small level of interdigitation between the lipids,
and considering a small area of the membrane mid-surface $dS = \left[ 1 + (\nabla h)^2/2 \right] d\v{r} + O(h^4)$ and curvature $c = \nabla^2 h + O(h^3)$,
the density fields are essentially uncoupled.
Hence, the internal elastic power of the membrane can be written as described in Ref.~\citep{Seifert1993}:
\begin{align}
	\m{H}_{int} = &\int_S dS
	\left[
		\frac{\sigma}{2} \left( \nabla h \right)^2 +
		\frac{\kappa}{2} \left( \nabla^2 h \right)^2 + \right. \nonumber \\
		&\left. \frac{k}{2} \left( \rho^+ + e \nabla^2 h \right)^2 + 
		\frac{k}{2} \left( \rho^- - e \nabla^2 h \right)^2
	\right],
  \label{e_source2}
\end{align}
where $\sigma$ is the membrane tension,
$\kappa$ is the membrane bending rigidity,
$k$ is the monolayer stretching coefficient,
and $e$ is the surface distance away from the membrane mid-surface [see Fig.~\ref{fig:model}].

The external elastic energy representing oscillatory loadings and the loading profile are defined as:
\begin{align}
	&\m{H}_{ext} = \int_{S_p} d\v{r} h p (\v{r}, t), 
	\label{energy_ext} \\
	&p (\v{r}, t) = p_0 \exp{\left( -\frac{12 |\v{r}|^2}{w^2} \right)} \sin{\left( \omega t\right)}, 
	\label{load_unit_surface}
\end{align}
where $p_0$ is the loading amplitude,
and $w$ is the width of the loading profile characterized by the Gaussian function.
The integration is performed in the area $S_p$, which is the projection onto the reference plane.
Considering the previous micropipette aspiration test in blood granulocytes~\citep{Evans1989},
where experiments were carried out with pipette sizes of $2$--$2.75$ $\mu$m and suction pressures of $\geq$ 1 Pa,
the loading amplitude $p_0$ and the width of the loading profile $w$ was set as $p_0 = 0.5$ Pa and $w = 1$ $\mu$m, respectively.
The loading area is equal to or smaller than the scan sizes in the AFM experiment ($\geq 2.0$ $\mu$m $\times$ $2.0$ $\mu$m)~\citep{Al-Rekabi2018}.
In {\it in vivo}, the shear stress varies from $1$ to $60$ dynes/cm$^2$ ($0.1$--$6$ Pa) depending in particular on vessel types~\citep{Loscalzo2002}.
Thus, the loading amplitude $p_0$ corresponds to the physiologically relevant stress in the microcirculation.
The scale of force applied to area $w$ corresponds to gaps in the endothelial barrier ($\sim 1$ $\mu$m) during the initial stages of transmigration of cancer cells~\citep{Chen2013},
and can be found, for example, in RBC-platelet~\citep{Takeishi2019JBSE} or RBC-microparticle~\citep{Takeishi2017} hydrodynamic interactions.
Representative snapshots of extending membranes are shown in Figs.~\ref{fig:lissajous}(a) and \ref{fig:lissajous}(b).
We also confirm that the given amplitude of oscillatory tensile loadings is small,
and hence the linear mechanical responses of the membrane to weak oscillations should be investigated, see panel (c) of Fig.~\ref{fig:lissajous}.

We have a dynamical equations of the membrane based on the Stokes approximation and the differentiation of the aforementioned Rayleighian~\eqref{R}.
The precise derivation of the equation is described in Appendix~\S\ref{secA1}.

\subsection{Dynamical equations and methodology}\label{subsec2_3}
The Fourier transforms of surface and bulk quantities in the $(x, y)$ plane are defined by:
\begin{align}
	  &f(\v{r}, t) = \int \frac{d \v{q}}{(2 \pi)^2} \hat{f} (\v{q}, t) e^{i \v{q} \cdot \v{r}}, 
	  \label{FFT_mem}
	  \\
	  &g(\v{r}, z, t) = \int \frac{d \v{q}}{(2 \pi)^2} \hat{g} (\v{q}, z, t) e^{i \v{q} \cdot \v{r}},
	  \label{FFT_bulk}
\end{align}
where $\v{q}$ is a wave-vector in the semi-Fourier space.
Thus, we have $\hat{h} (\v{q}, t)$ and $\hat{\rho} (\v{q}, t)$ by Eq.~\eqref{FFT_mem},
and $\hat{\v{V}} (\v{q}, z, t)$ by Eq.~\eqref{FFT_bulk}.
The caret symbol ``$\hat{\,}$'' is omitted below (e.g., $\hat{h}(\v{q}, t)$ is simply rewritten as $h (\v{q}, t)$).

Assuming axial symmetry with respect to the $z$-axis ($x=y=0$),
we obtain the linear time-evolution equation of $h(\v{q}, t)$ and $\rho (\v{q}, t)$ as:
\begin{align}
\partial_t
\begin{pmatrix}
	q h \\
	q \rho
\end{pmatrix}
&= -\mathbf{M}(q)
\begin{pmatrix}
	q h (\v{q}, t) \\
	q \rho (\v{q}, t)
\end{pmatrix}
\label{governing_e1}
\\ \nonumber
&+
\begin{pmatrix}
	\dfrac{\pi p_0 w^2}{48 \eta} \exp{\left(- w^2 q^2/48 \right)} \sin{\left( \omega t\right)} \\
	0
\end{pmatrix},
\end{align}
where $\mathbf{M}(q)$ is the dynamical matrix:
\begin{equation}
\mathbf{M}(q) = 
\begin{pmatrix}
	\dfrac{\sigma q + \tilde{\kappa} q^3}{4 \eta} & -\dfrac{k e q}{4 \eta} \\
	-\dfrac{k e q^4}{b + \eta q + \eta_s q^2} & \dfrac{k q^2}{2 \left( b + \eta q + \eta_s q^2 \right)}
\end{pmatrix}.
	\label{governing_e2}
\end{equation}
By reference to previous experimental works~\citep{Bitbol2012, Merkel1989, Rawicz2000, Pott2002, Horner2013},
the following standard parameter values for lipid bilayers~\citep{Fournier2015} are used in this study:
$\tilde{\kappa}$ ($= \kappa + 2 k e^2$)
is the effective bending rigidity at fixed lipid densities~\citep{Seifert1993},
$\kappa = 10^{-19}$ J,
$k =  0.1$ J/m$^2$,
$e = 1.0$ nm,
$b = 10^9$ J$\cdot$s/m$^4$,
$\eta = 10^{-3}$ J$\cdot$s/m$^3$,
and $\eta_s$ ($= \eta_2 + \lambda_2/2$) is the surface viscosity ($= 10^{-9}$ J$\cdot$s/m$^2$)~\citep{Bitbol2012, Merkel1989, Rawicz2000, Pott2002, Horner2013}.
For instance, those model parameters about human RBCs were reviewed in~\citep{Tomaiuolo2014}:
the surface viscosity ($\eta_s$) is $(7 \pm 2) \times 10^{-7}$ J$\cdot$s/m$^2$,
the intermonolayer friction coefficient ($b$) is $0.2$--$1.2 \times 10^9$ J$\cdot$s/m$^4$,
the bending elastic modulus ($\kappa$) is $(1.15 \pm 0.9) \times 10^{-19}$ J,
the area compressibility modulus corresponding to the monolayer stretching coefficient ($k$) is $0.399 \pm 0.1$1 J/m$^2$.
Since the surface viscosity of the RBC is influenced by structure and the mechanical integrity of the spectrin network (RBC skeleton),
the resultant value tends to be greater than pure lipid bilayers.
Overall, order of the magnitude of model parameters of lipid bilayers in this study are consistent with those in human RBC.

The explicit fourth-order Runge-Kutta method is used for the time integration.
Owing to the axial symmetry assumption,
the Fourier transform leads to Hankel transform,
and we have 
\begin{align}
\begin{pmatrix}
	h (r, t) \\
	\rho (r, t) \\
	\tau_{zz} (r, t)
\end{pmatrix}
= \frac{1}{2 \pi} \int_0^\infty dq
\begin{pmatrix}
	h (q, t) \\
	\rho (q, t) \\
	\tau_{zz} (q, t)
\end{pmatrix}
\m{J}_0 (qr) q,
\label{hankel_transform}
\end{align}
where $\tau_{zz}$ is the normal stress acting on the membrane, and $\m{J}_0$ is the Bessel function of the first kind.
In particular, the form of $\tau_{zz}^\pm$ in Fourier space is:
\begin{align}
	\tau_{zz}^\pm (\v{q}) = 
	&\mp 2 \eta q \partial_t h(\v{q}) \nonumber \\
	&\pm \frac{\pi p_0 w^2}{24} \exp{\left( - w^2 q^2/48 \right)} \sin{\left( \omega t\right)}.
	\label{normal_stress}
\end{align}
The first term on the right-hand side represents the ($z$, $z$) component of the liquid stress tensor $T_{ij} \equiv -P\delta_{ij} + \eta \left( \partial_i v_j + \partial_j v_i \right)$,
evaluated for the upper and lower monolayers;
$P$ is a one of the Lagrange multiplier fields (see Appendix~\S\ref{secA1}) corresponding to the hydrostatic pressure;
and $\delta_{ij}$ is the Kronecker delta.

\subsection{Analysis of dynamical viscoelasticity}\label{subsec2_4}
We assume a linear mechanical response of the membrane to weak oscillatory strains $\varepsilon (t) = \varepsilon_0 \exp{(i \omega t)}$,
and evaluate the stress on the membrane $\Sigma (t) = \Sigma_0 \exp{i (\omega t + \delta)}$ as:
\begin{equation}
	\Sigma (t) = E^\ast (\omega) \varepsilon (t),
	\label{linear_response}
\end{equation}
where $\delta$ is the phase difference between the tensile strain $\varepsilon (t)$ and stress $\sigma (t)$.
$E^\ast (\omega)$ can be decomposed into two components:
\begin{align}
	E^\ast (\omega)
	= \frac{\Sigma (t)}{\varepsilon (t)}
	= \frac{\Sigma_0}{\varepsilon_0} \left( \cos\delta + i \sin\delta \right)
	= E^\prime (\omega) + i E^{\prime\prime} (\omega),
	\label{E}
\end{align}
where the real part $E^\prime (\omega) (= (\Sigma_0/\varepsilon_0) \cos\delta)$ is the storage modulus representing the elastic component of the stress,
and the imaginary part $E^{\prime\prime} (\omega) (= (\Sigma_0/\varepsilon_0) \sin\delta)$ is the loss modulus representing the viscous part~\citep{Mewis2012, Squires2010}.
We also define the loss tangent as
\begin{equation}
	\tan{\delta} = E^{\prime\prime}/E^\prime.
\end{equation}
Although various experimental techniques have been proposed to measure the dynamical viscosity of the membrane,
it is still a challenge to track the lipid dynamics during deformation.
Optical tweezer experiments showed that a membrane shape corresponding to the membrane strain of $1$,$2$-dioleoyl-$sn$-glycero-$3$-phosphocholine (DOPC) giant unilamellar vesicles reflects applied forces~\citep{Lee2008}.
In this study, instead of introducing membrane area strain,
we can directly track the lipid molecular density,
which is the basis for dynamical models including lipid tilt near molecular inclusions or physiochemical interaction of lipids.
In this study, therefore, the strain $\varepsilon (t)$ and stress $\Sigma(t)$ are evaluated by the scaled mass density in upper monolayer $\rho^+ (t)$ in Eq.~\eqref{strain_zz} [or the mean scaled mass density, $\rho = \left(\rho^+ - \rho^- \right)/2$] and the normal stress acting on the membrane in upper monolayer $\tau_{zz}^+ (t)$ in Eq.~\eqref{normal_stress}.
Since the applied loading amplitude $p_0$ is still small,
and since it is practically difficult to quantify the dynamical viscoelasticity of the whole membrane,
the aforementioned scaled density and normal stress are evaluated at the center ($r = 0$) of the membrane (i.e., $\rho^+|_{r = 0}$ and $\tau_{zz}^+|_{r = 0}$),
where the amplitudes are maximized.

\section{Results}\label{sec3}

\begin{figure}[htbp]
  \centering
    \includegraphics[clip,width=7.5cm]{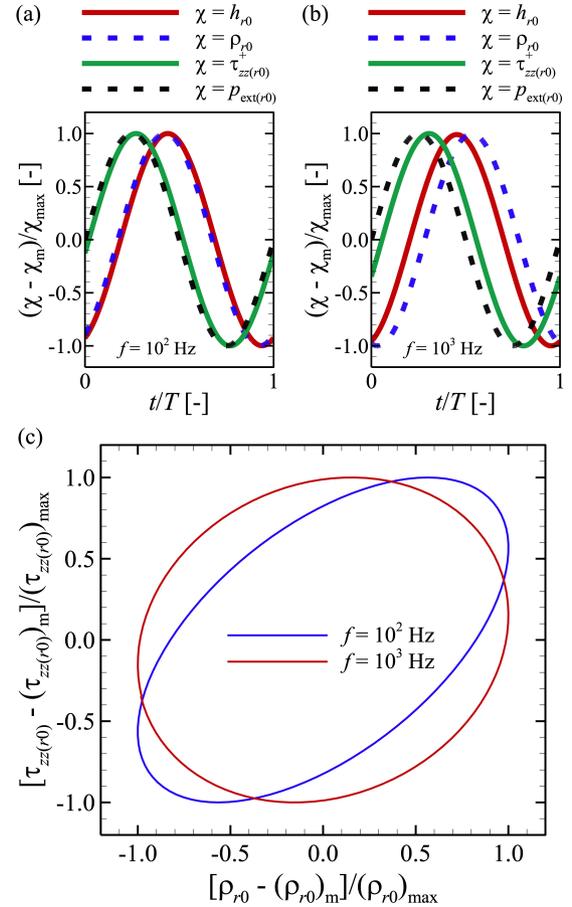}
    \caption{
    	    (a, b) Time history of the calculated values $\chi$:
	    membrane height $h_{r0}$ (red solid line),
	    mean scaled density $\rho_{r0}$ (blue dashed line),
	    membrane stress $\tau_{zz(r0)}^+$ (green solid line),
	    and external load $p_{\mathrm{ext}(r0)}$ (black dashed line) at the center of the membrane $r = 0$ during a period $T$,
	    obtained with (a) $f = 10^2$ Hz and (b) $f = 10^3$ Hz.
	    These values are normalized by each amplitude $\chi_\mathrm{max}$,
	    and shifted so that each baseline is the mean value $\chi_{\mathrm{m}}$.
	    (c) Lissajous-Bowditch plots of normalized membrane stress $\left[ \tau_{zz(r0)}^+ - (\tau_{zz(r0)}^+)_\mathrm{m} \right]/(\tau_{zz(r0)}^+)_\mathrm{max}$ versus normalized scaled mass density $\left[ \rho_{r0} - (\rho_{r0})_\mathrm{m} \right]/(\rho_{r0})_\mathrm{max}$ for different $f$ ($= 10^2$ and $10^3$ Hz).
    	    The results are obtained with infinitesimally small surface tension $\sigma = 0$.
    }
    \label{fig:lissajous}
\end{figure}

First, we investigate the membrane behavior under oscillatory tensile loadings with a specific frequency $f$ for infinitesimally small surface tension $\sigma = 0$ N/m.
Figures~\ref{fig:lissajous}(a) and \ref{fig:lissajous}(b) show one cycle of $h$,
the mean scaled density $\rho$,
i.e., $\rho = \left(\rho^+ - \rho^- \right)/2$,
corresponding to the upper monolayer scaled density $\rho^+$,
the normal stress on the upper monolayer along the $z$-direction $\tau_{zz}^+$,
and external loads $p_\mathrm{ext}$ at the center of the membrane ($r = 0$) for different loading frequencies $f$ ($= 10^2$ Hz and $10^3$ Hz) after they have fully developed,
where these values are normalized by each amplitude $\chi_\mathrm{max}$,
and shifted so that each baseline is the mean value $\chi_{\mathrm{m}}$.
Since the membrane height $h$ and density $\rho$ at $r = 0$ (i.e., $h|_{r = 0}$ and $\rho|_{r = 0}$) drift as time passes,
we use data after they have fully saturated to quantify the phase differences between these values.
Hereafter, we redescribe $h|_{r=0} \to h_{r0}$, $\rho|_{r = 0} \to \rho_{r0}$, $\tau_{zz}^+|_{r = 0} \to \tau_{zz(r0)}^+$, and $p_\mathrm{ext}|_{r = 0} \to p_{\mathrm{ext}(r0)}$.
At relatively low frequency $f = 10^2$ Hz,
there is no significant phase differences between $\tau_{zz(r0)}^+$ and $p_{\mathrm{ext}(r0)}$ and between $h_{r0}$ and $\rho_{r0}$,
where the latter values ($h_{r0}$ and $\rho_{r0}$) are delayed from the other two ($\tau_{zz(r0)}^+$ and $p_{\mathrm{ext}(r0)}$) [Fig.~\ref{fig:lissajous}(a)].
A higher frequency condition ($f = 10^3$ Hz) causes apparent phase differences between these values,
where the stress $\tau_{zz(r0)}^+$ and scaled density $\rho_{r0}$ start to be later for the external load $p_{\mathrm{ext}(r0)}$ and height $h_{r0}$, respectively [Fig.~\ref{fig:lissajous}(b)].

We also investigate Lissajous-Bowditch plots~\citep{Mujumdar2002, Lee2003} of $\tau_{zz(r0)}^+$ versus $\rho_{r0}$,
following the classical method of analyzing nonlinear oscillations.
We recall that an elastic solid corresponds to a straight line deviated at $\pi/4$ from the $y$-axis,
a Newtonian fluid corresponds to a perfect circle,
and viscoelastic materials under small-amplitude oscillatory loadings correspond to an ellipse whose major axis deviates at $\pi/4$ from $y$-axis.
Nonlinear responses can generally lead to a more complex shape,
which itself is a characterization of the nonlinear viscoelasticity of the suspension~\citep{Mujumdar2002},
as is usually observed under large-amplitude oscillatory loadings.
The results for the membrane are shown in Fig.~\ref{fig:lissajous}(c),
where the Lissajous plots are asymptotic from the ellipsoidal shape to the circle shape as $f$ increases.
The result indicates that the given amplitude is small enough to assume a linear mechanical response of the membrane to oscillatory tensile loadings.
Hereafter, we consider small-amplitude oscillatory loadings.

\begin{figure*}[htbp]
  \centering
    \includegraphics[clip,width=13cm]{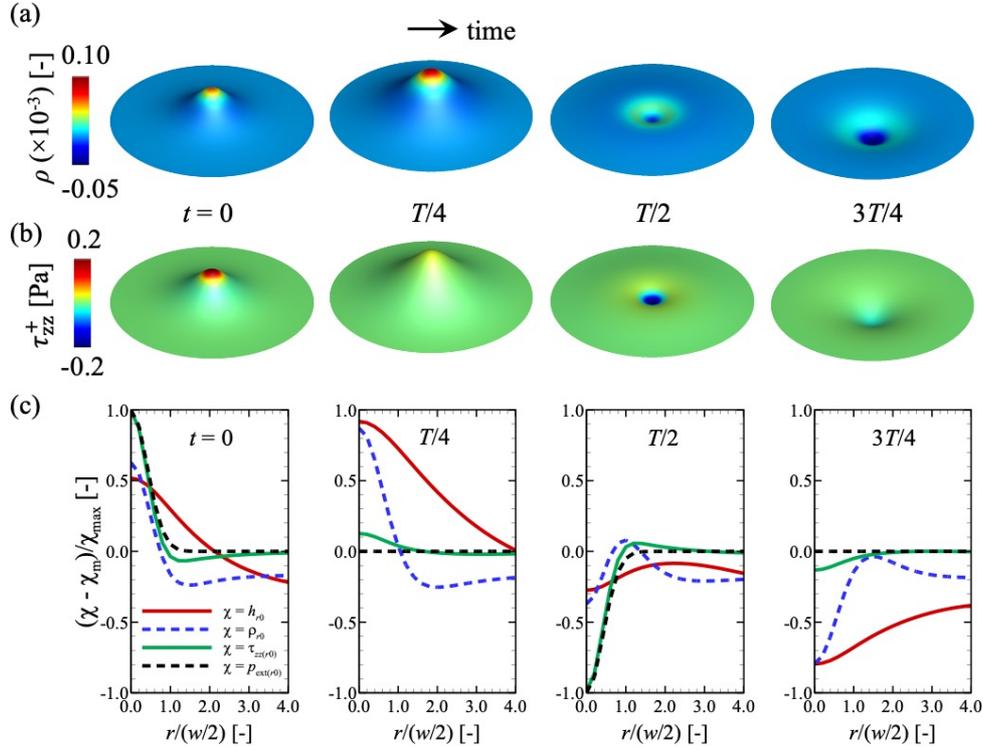}
    \caption{
           (a, b) Representative snapshots of an extending membrane under tensile loadings for $f = 10^2$ Hz,
	    where color contours represent (a) the mean scaled density $\rho$ and (b) the membrane stress $\tau_{zz}^+$.
	    (c) Distribution of the 
	    calculated values $\chi$ at each time period ($t = 0$, $T/4$, $T/2$, and $3T/4$) for $f = 10^2$ Hz.
	    The results are obtained with $\sigma = 0$.
    }
    \label{fig:time_hist}
\end{figure*}

Figures~\ref{fig:time_hist}(a) and \ref{fig:time_hist}(b) show a series of representative snapshots of the deformed membrane during loading cycle $T$ ($= 1/f = 10^{-2}$ s),
where color contours represent the mean scaled density $\rho$,
and the stress in the tensile direction $\tau_{zz}^+$.
At the phase when the height of the membrane ($h_{r0}$) is the maximum ($t = 0$),
$\tau_{zz(r0)}^+$ is large,
while the density responds late [Figs.~\ref{fig:time_hist}(a) and \ref{fig:time_hist}(b)].
Figure~\ref{fig:time_hist}(c) shows one cycle of $h_{r0}$, $\rho_{r0}$, $\tau_{zz(r0)}^+$, and $p_{\mathrm{ext}(r0)}$ after they have fully developed ($t \geq 1$ s).

\begin{figure}[htbp]
  \centering
  \includegraphics[clip,width=6cm]{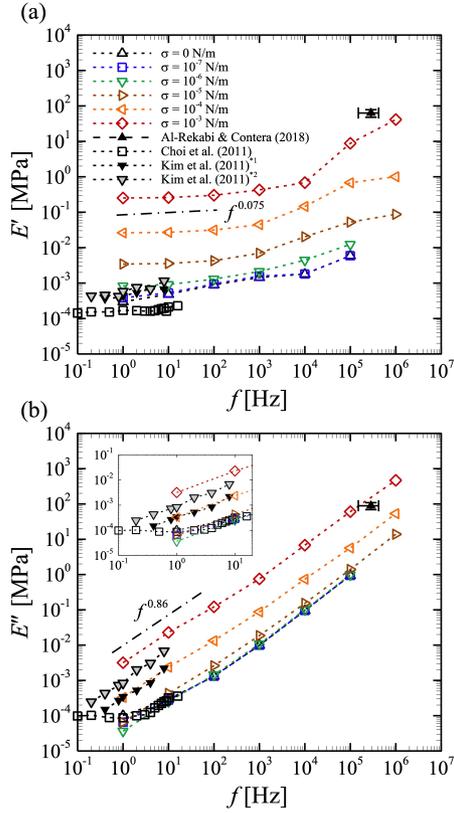}
  \caption{
	    Storage modulus (a) $E^\prime$ and loss modulus (b) $E^{\prime\prime}$ as a function of frequency $f$ for different membrane tensions $\sigma$.
	    Experimental data by \cite{Al-Rekabi2018, Choi2011, Kim2011} are also plotted,
	    where the data of~\citep{Kim2011} at the oscillatory amplitude of $8$ mN/m$^{(\ast 1)}$ and $10$ mN/m$^{(\ast 2)}$ are shown.
	    Power laws obtained with the largest $\sigma$ ($= 1$ mN/m) for $E^\prime$ in relatively small $f \leq 10^3$ Hz and for $E^{\prime\prime}$ in the whole range of $f$ are also shown as dash-dot lines.
  }
  \label{fig:E}
\end{figure}

Considering Eq.~\eqref{E},
the storage modulus $E^\prime$ and loss modulus $E^{\prime\prime}$ can be quantified using $\rho^+_{r0}$ and $\tau_{zz(r0)}^+$,
and the results are shown in Figs.~\ref{fig:E}(a) and \ref{fig:E}(b), respectively.
Previous AFM experimental data of DPPC bilayers by \cite{Al-Rekabi2018} are also plotted,
where $E^\prime = 63 \pm 15.3$ MPa and $E^{\prime\prime} = 89 \pm 20.0$ MPa for high frequencies $f = 150$--$420$ kHz.
Furthermore, experimental data of liquid-condensed (LC-)DPPC monolayers for low frequencies $O(f) = 10^{-1}$--$10^1$ Hz by \cite{Choi2011, Kim2011} are also plotted,
where the surface viscoelasticity $E_s^\ast$ were quantified by applying oscillatory magnetic field.
Assuming 1-nm thickness of LC-DPPC,
corresponding to the half of the bilayer thickness in our model (i.e., $e = 1$ nm),
the surface viscoelasticity was converted to our metrics, i.e., $E^\ast = E_s^\ast/e$~\citep{Shkulipa2005}.
A more precise description about the translational and rotational drug on a cylinder in a membrane should be referred to~\citep{Hughes1981}.
Our numerical results show that both the storage modulus $E^\prime$ and loss modulus $E^{\prime\prime}$ increase with frequency $f$ (Fig.~\ref{fig:E}).
These tendencies are consistent with those in previous experimental results for low frequencies~\citep{Choi2011, Kim2011}.
Especially at the highest membrane tension $\sigma = 1$ mN/m,
both values agree well with previous AFM measurements~\citep{Al-Rekabi2018}.
We evaluate the calculated $E^\prime$ and $E^{\prime\prime}$ based on the power law of $y = \beta x^\alpha$.
At the highest $\sigma$ ($= 1$ mN/m), for instance, the increases of $E^\prime$  for relatively low $f$ ($\leq 10^3$ Hz) can be well approximated as $E^\prime \propto f^{0.075}$,
while $E^{\prime\prime}$ for the whole range of $f$ ($10^0$ Hz $\leq f \leq$ $10^6$ Hz) that we investigated can be well approximated as $E^{\prime\prime} \propto f^{0.86}$.
Note that the mean value and standard deviation of $\alpha$ for different $\sigma$ were $0.103 \pm 0.059$ Pa in $E^\prime$ and $0.753 \pm 0.251$ Pa in $E^{\prime\prime}$.
These results indicate that a lipid bilayer membrane characterized by a fluid membrane cannot be modeled as Maxwell materials (Appendix~\S\ref{secA2}),
where $E^\prime \propto f^2$ and $E^{\prime\prime} \propto f$ for low $f$.
\begin{figure}[htbp]
  \centering
    \includegraphics[clip,width=6cm]{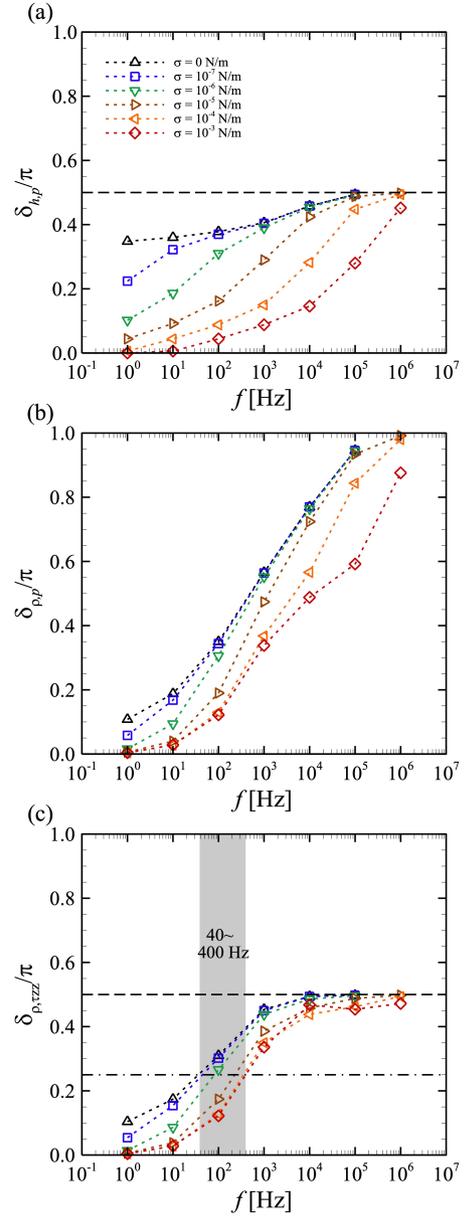}
    \caption{
    	    Phase differences (a) $\delta_{h,p}$ between $h_{r0}$ and $p_{\mathrm{ext}(r0)}$,
	    (b) $\delta_{\rho,p}$ between $\rho_{r0}$ and $p_{\mathrm{ext}(r0)}$, and 
	    (c) $\delta_{\rho,\tau_{zz}^+}$ between $\tau_{zz(r0)}^+$ and $\rho_{r0}$ as a function of frequency $f$ for different surface tensions $\sigma$.
	    Dashed and dash-dot lines in panel (a) and (c) represent specific phase difference values of $0.5$ and $0.25$, respectively.
    }
    \label{fig:loss_tan}
\end{figure}
\begin{figure}[htbp]
   \centering
   \includegraphics[clip,width=6cm]{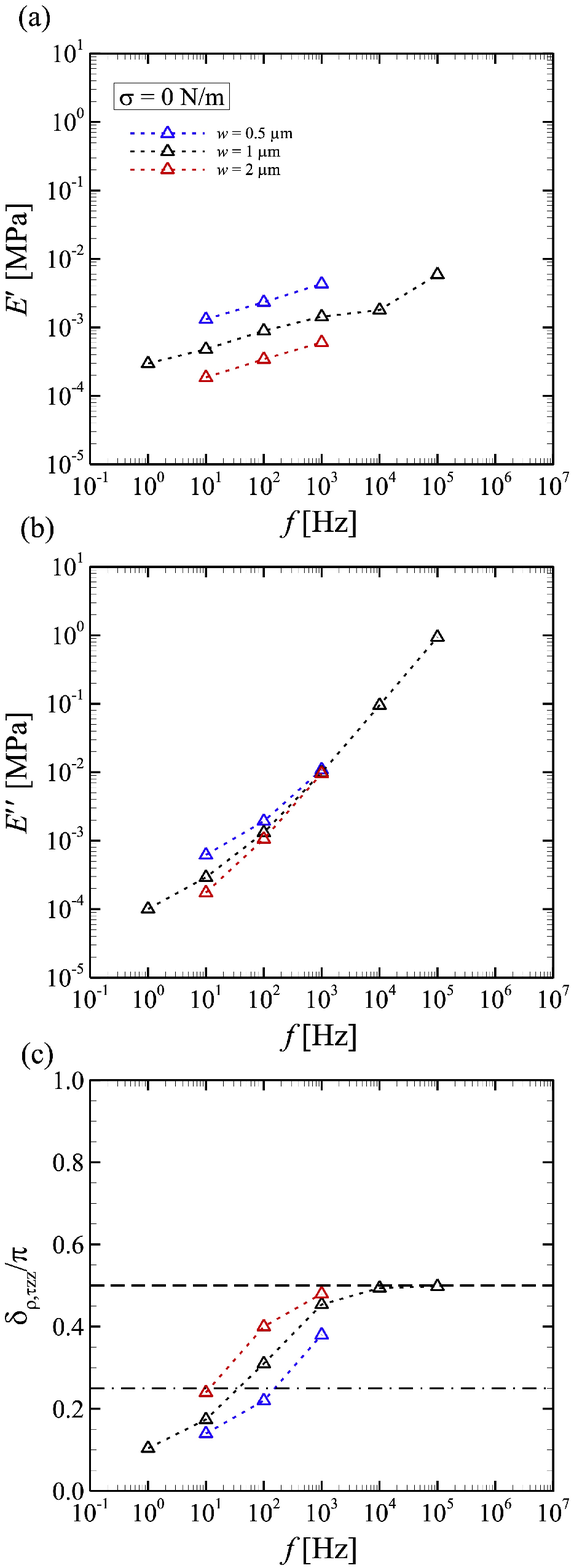}
   \caption{
    	    (a) $E^\prime$,
	    (b) $E^{\prime\prime}$, and
	    (c) $\delta_{\rho,\tau_{zz}^+}$ at $\sigma = 0$ in $10^1 \leq f \leq 10^3$ Hz for different widths of loading profile $w$ ($= 0.5$, and $2$ $\mu$m).
    }
    \label{fig:comparison}
\end{figure}

To investigate the effects of frequency $f$ and membrane tension $\sigma$,
the phase differences between $h_{r0}$,
$\rho_{r0}$,
$\tau_{zz(r0)}^+$, and 
$p_{\mathrm{ext}(r0)}$ as a function of frequency $f$ are shown in Figs.~\ref{fig:loss_tan}(a)--(c).
The phase difference $\delta_{h, p}$ between $h_{r0}$ and $p_{\mathrm{ext}(r0)}$ is shown in Fig.~\ref{fig:loss_tan}(a).
The results show that $\delta_{h, p}$ gradually reaches $\pi/2$ as the frequency $f$ increases and as the membrane tension $\sigma$ decreases [Fig.~\ref{fig:loss_tan}(a)].

On the other hand,
the phase difference $\delta_{\rho, p}$ between $\rho_{r0}$ and $p_{\mathrm{ext}(r0)}$ gradually reaches $\pi$ as the frequency $f$ increases and as the membrane tension $\sigma$ decreases [Fig.~\ref{fig:loss_tan}(b)].
In comparison with $\delta_{h, p}$ [Fig.~\ref{fig:loss_tan}(a)],
the effect of the membrane tension $\sigma$ on the phase difference $\delta_{\rho, p}$ is relatively small.

Similar to $\delta_{h, p}$,
the phase difference $\delta_{\rho, \tau_{zz}^+}$ between $\tau_{zz(r0)}^+$ and $\rho_{r0}$ reaches $\pi/2$ as the frequency $f$ increases and as the membrane tension $\sigma$ decreases [Fig.~\ref{fig:loss_tan}(c)].
Especially for higher frequencies $f \geq 10^4$ Hz,
$\delta_{\rho, \tau_{zz}^+}$ are uniformly close to $\pi/2$ independently of $\sigma$ [Fig.~\ref{fig:loss_tan}(c)].

Considering the specific value of $\delta_{\rho, \tau_{zz}^+}$ = $\pi/4$,
i.e., the loss tangent $\tan{(\delta_{\rho, \tau_{zz}^+})} = 1$,
the membrane characteristic shifts from elastic-dominant [i.e., $\delta_{\rho, \tau_{zz}^+} <$ $\pi/4$ and $\tan{(\delta_{\rho, \tau_{zz}^+})} <$ 1] to viscous dominant [i.e., $\delta_{\rho, \tau_{zz}^+} >$ $\pi/4$ and $\tan{(\delta_{\rho, \tau_{zz}^+})} >$ 1] as the frequency $f$ increases [Fig.~\ref{fig:loss_tan}(c)].
Furthermore,
the transition appears within the range of $40$ Hz $\leq f \leq$ $400$ Hz for all $\sigma$ that we investigated [Fig.~\ref{fig:loss_tan}(c)].

We also investigated the effect of the width of the loading profile $w$ on viscoelastic metrics.
The results of $E^\prime$, $E^{\prime\prime}$, and $\delta_{\rho,\tau_{zz}^+}$ at $\sigma = 0$ for $10^1$ Hz $\leq f \leq$ $10^3$ Hz,
obtained with different widths of the loading profile $w$ ($= 0.5$ and $2$ $\mu$m) are superposed on Figs.~\ref{fig:E} and \ref{fig:loss_tan}(c).
The results are shown in Fig.~\ref{fig:comparison}.
The storage modulus $E^\prime$ tends to increase as $w$ decreases [Fig.~\ref{fig:comparison}(a)], while $E^{\prime\prime}$ remains almost the same [Fig.~\ref{fig:comparison}(b)].
As a consequence, the phase difference $\delta_{\rho,\tau_{zz}^+}$ increases with $w$ [Fig.~\ref{fig:comparison}(c)].
A small $w$ leads to a large curvature,
resulting in large internal elastic energy (or power) of the membrane and large $E^\prime$.
The results indicate that apparent fluidity of a $2$D fluid membrane becomes greater with larger loading profile $w$.

\section{Discussion}\label{sec4}
Recent experimental techniques have made it possible to measure the dynamical viscoelasticity of the lipid bilayer,
where classical viscoelastic metrics as complex ``shear'' moduli, $G^\ast (\omega) = G^\prime (\omega) + i G^{\prime\prime} (\omega)$, are quantified assuming a linear mechanical response of the membrane to oscillatory shear strain~\citep{Choi2011, Harland2011, Kim2011}.
These attempts have shown that monolayers of LC-DPPC tend to behave as viscous dominant ($G^\prime < G^{\prime\prime}$) as frequency increases,
especially for $O(f) \geq 10^0$ Hz~\citep{Choi2011, Kim2011}.
It is also known that complex shear moduli of a phospholipid bilayer composed of $1$,$2$-dimyristoyl-sn-glycero-$3$-phospho-choline (DMPC) are affected by membrane states or temperatures,
wherein a viscous-dominant ($G^{\prime\prime} > G^\prime$) state is uniformly present at the liquid-gel transition temperature ($= 23.5$ $^\circ$C) independently of frequency,
while at a liquid phase temperature ($= 20.1$ $^\circ$C) or gel phase temperature ($= 25.8$ $^\circ$C),
the viscous-dominant state occurs at low frequencies ($O(f) < 10^0$ Hz) and the elastic-dominant ($G^{\prime\prime} < G^\prime$) state occurs at high frequencies ($O(f) > 10^0$ Hz)~\citep{Harland2011}.
Applying tensile loadings using AFM,
Al-Rekabi et al. showed the effect of cholesterol concentration on the dynamical viscoelasticity~\citep{Al-Rekabi2018},
at relatively high frequencies ($\sim 150$--$420$ kHz),
of a lipid bilayer composed of a DPPC-cholesterol mixture.
Despite these insights,
the dynamical viscoelastic characteristics of a bilayer composed of pure phospholipid (e.g., DPPC) has not yet been fully described,
especially under tensile loadings.
Furthermore, there is still no consensus on the transition mechanism between the viscous- and elastic-dominant states in lipid bilayers.
We therefore tackle this issue by model analysis following a previous theoretical and numerical study by \cite{Fournier2015},
and quantify the dynamical viscoelasticity of an almost planar bilayer membrane, characterized as a $2$D fluid membrane, under oscillatory tensile loadings.
We obtain hydrodynamical equations of bilayer membranes using Onsager's variational principle,
which is an established unified framework for the dissipative dynamics of soft matter systems~\citep{Doi2011, Goldstein2001}.

In this study, viscoelastic metrics $E^\ast (\omega) = E^\prime (\omega) + i E^{\prime\prime} (\omega)$ as a response to tensile loadings are introduced,
and are quantified by the time difference between the scaled mass density $\rho$ and normal stress acting on the membrane $\tau_{zz}$.
Our numerical results show that membrane characteristics shift from the elastic-dominant state to the viscous-dominant state ($E^{\prime\prime} > E^\prime$) when the frequency $f$ increases (Fig.~\ref{fig:loss_tan}),
which is consistent with previous AFM experimental measurements in DPPC bilayers~\citep{Al-Rekabi2018}
and in cells~\citep{Rigato2017}.
The resultant viscoelastic behavior of the membrane under tensile loadings cannot be estimated by the well-known Maxwell materials,
where complex moduli can be estimated as $E^\prime \propto f^2$ and $E^{\prime\prime} \propto f$ for low $f$ (see also Appendix~\S\ref{secA2}).
Our study further provides conclusive evidence that a $2$D fluid membrane featuring lipid bilayers exhibits a transition between the viscous- and elastic-dominant states depending on oscillatory loading frequencies.
Such transitions appear within the range of $40$ Hz $\leq f \leq$ $400$ Hz at a 1-$\mu$m width ($w = 1$ $\mu$m) of the loading profile for all surface tensions $\sigma$ that we investigated ($0 \leq \sigma \leq 1$ mN/m).
The transition shifts to a lower frequency range as $w$ increases (Fig.~\ref{fig:comparison}).
These numerical results,
especially at $w = 1$ $\mu$m,
suggest that vesicle membranes behave almost as elastic sheets under a physiological human heartbeat,
which is close to $1$ Hz,
and that viscous characteristics emerge at high frequencies $O(f) \geq 10^1$ Hz,
which may occur with artificial blood pumps.
Therefore, this knowledge may be helpful for designing novel artificial blood pumps to prevent the risk of vesicle rupture.
More recently, the use of microbubbles coated with a biocompatible shell (e.g., lipid bilayers) as an ultrasound contrast agent has attracted attention not only for echocardiography but also for tumor detection and other therapeutic purposes~\citep{Versluis2020}.
The range of microbubble sizes appropriate for clinical use corresponds to resonance frequencies on the order of $1$--$10$ MHz.
However, the dynamical viscoelasticity of such a shell is still debated.
Our numerical results provide insights into its design,
in terms of the mechanical properties necessary for adequate adaption to surrounding solvents with high oscillatory frequencies.

\begin{figure}[htbp]
  \centering
    \includegraphics[clip,width=7cm]{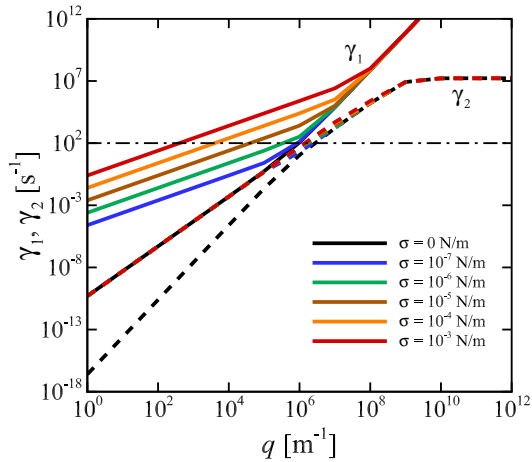}
    \caption{
	    The eigenvalues of the dynamical matrix $\mathbf{M}(q)$ in Eq.~\eqref{governing_e2} as a function of wave-vector $q$ for different surface tensions $\sigma$: $\gamma_1 (q)$ (solid lines) and $\gamma_2 (q)$ (dashed lines).
	    The results are obtained with the standard values given in the text.
	    The dash-dot line represents the eigenvalues at $100$ Hz.
    }
    \label{fig:eigen}
\end{figure}

To understand the transition mechanism between the viscous- and elastic-dominant states,
we calculate the relaxation rates,
which correspond to the eigenvalues $\gamma_i (q)$ ($i = 1$ or $2$) of the dynamical matrix $\mathbf{M}(q)$ in Eq.~\eqref{governing_e2} as a function of wavenumber $q$ for different surface tensions as shown in Fig.~\ref{fig:eigen},
where $\gamma_i (q)$ are obtained with the standard values given below Eq.~\eqref{governing_e2}.
The values of $\gamma_1$ increase with $\sigma$ at relatively low $q$ and collapse on a single curve for high $q$.
The values of $\gamma_2$ are much lower than those of $\gamma_1$,
and the two follow almost the same line except for infinitesimally small $\sigma$ ($= 0$).
It is known that $\gamma_1$ corresponds to the relaxation of $h$ at fixed $\rho$,
while $\gamma_2$ corresponds to the relaxation of $\rho$ at fixed $h$~\citep{Fournier2015}.
Considering the scale of loading area $w$ ($= 1$ $\mu$m),
let us take a representative wavelength $\lambda$ ($= 2 \pi/q$) $= 1$ $\mu$m,
corresponding to $q \geq 10^6$ m$^{-1}$.
The values of $\gamma_1$ collapse for over $10^7$ Hz,
while those of $\gamma_2$ collapse around $10^2$ Hz independently of $\sigma$.
For lower loading frequencies $f < 10^2$ Hz,
$\gamma_i$ for $q >10^6 \approx 1/w$ are higher than $f$.
This indicates that the membrane exhibits fast relaxation or immediately respond to loadings,
resulting in small phase differences.
For higher loading frequencies $f > 10^2$ Hz,
$\gamma_i$ lower than $f$ appears for $q > 1/w$.
Hence, the membrane cannot fully attenuate the scaled density $\rho$,
whose relaxation is represented by $\gamma_2$,
under such high loading frequencies.
Consequently, the phase difference $\delta_{\rho,\tau_{zz}^+}$ increases.
\citet{Rigato2017} experimentally studied the rheological behavior of the mouse fibroblast-like cell line $3$T$3$ in a vast frequency range by treating the cells with four different drugs.
They quantified viscoelastic properties with frequency-dependent complex shear moduli ($G^\prime$ and $G^{\prime\prime}$).
Although the magnitudes of the modulus were different from those of the metrics in our study ($E^\prime$ and $E^{\prime\prime}$),
they also exhibited state transitions at specific frequencies.
Furthermore, the transition frequency was lower in cells with disrupted actin or reduced prestress ($28$ kHz and $56$ kHz, respectively; compared to $84$ kHz for untreated cells)~\citep{Rigato2017}.
These experimental results are consistent with our numerical results obtained with larger surface tensions $\sigma$ [Fig.~\ref{fig:loss_tan}(c)].
A more recent theoretical study by \cite{Hang2022} proposed a hierarchical model,
which is in broad agreement with all existing experimentally measured $G^\prime$ and $G^{\prime\prime}$ in~\citep{Rigato2017}.
Despite these efforts,
the authors acknowledged that there is still no consensus on the mechanism of the change.
Our numerical results of $\delta_{\rho,\tau_{zz}^+}$ [Fig.~\ref{fig:loss_tan}($c$)] and the relaxation rates (Fig.~\ref{fig:eigen}) may provide one explanation for the problem.
Although some similarities are found in our metrics $E^\ast$ and complex shear moduli $G^\ast$,
rigorous experimental measurements of the delay of resultant membrane strain (or force) from given forces (or membrane strain) are required to relate with each other,
which will be the topic of a future study.

The calculated complex moduli ($E^\prime$ and $E^{\prime\prime}$) obtained with the specific surface tension $\sigma = 1$ mN/m agree well with those in DPPC bilayers~\citep{Al-Rekabi2018}.
Although surface tension depends on various conditions,
e.g., temperature and cholesterol concentration,
the orders of magnitude that we investigated ($0 \leq \sigma \leq 1$ mN/m) cover the physiologically relevant surface tension not only of a bilayer composed of pure phospholipid
but also of some cell membranes such as lined structures.
For instance,
the interfacial tension of a lipid bilayer was numerically estimated in the range of $6$ to $200$ mN/m~\citep{Feller1996, Jahnig1996}.
By applying a pressure difference on both sides of the membrane and measuring its curvature,
the interfacial tension of lipid bilayers is identified as $3.4 \pm 0.6$ mN/m,
with variation depending on the electrolyte composition~\citep{Coster1968}:
$1.623$ mN/m and $4.715$ mN/m for pure lecithin and pure cholesterol membranes, respectively~\citep{Petelska1998}.
Using AFM,
the lateral tension of pore-spanning lipid bilayers composed of N,N,-dimethyl-N,N,-dioctadecylammonium bromide (DODAB) was estimated as $1.0$ mN/m in the fluid state,
and as $5.0$ mN/m in the gel state~\citep{Steltenkamp2006}.
Considering these studies,
the order of magnitude of surface tension of lipid bilayers in the fluid state can be estimated as $O(\sigma) = 1$ mN/m.
Hence, our numerical results for complex moduli and the phase differences obtained with $\sigma = 1$ mN/m potentially represent the dynamical viscoelasticity of a lipid bilayer membrane in the fluid state and at fixed temperature.

Although we considered a homogeneous lipid bilayer membrane,
cell membranes {\it in vivo} contain non-negligible numbers of various proteins,
e.g., cholesterol and phosphatidylcholine,
and are embedded with various functional molecules such as peptides, proteins, and polysaccharides.
Previous studies have revealed that changes to the local mechanical properties of the cell membrane regulate the propagation of forces in cells~\citep{Janmey2006}.
Moreover, it has been reported that mechanical properties can modulate a membrane's interface with its surrounding liquid and selectively control ionic adsorption and condensation~\citep{Contera2010}.
Hence, it would be interesting to study how these factors,
which may be roughly represented by stretching or bending stiffness in our membrane model,
change the dynamical viscoelasticity of the membrane.
Further systematic analysis to investigate the effect of membrane parameters on $E^\prime$ and $E^{\prime\prime}$ will constitute a future study.

Our numerical results and quantitative model analysis of the dynamical viscoelasticity of lipid bilayers characterized as a $2$D fluid membrane will be helpful to build more rigorous continuum membrane models that consider multi-scale dynamics,
and to gain insights not only into active or passive cell dynamics 
but also into vesicle design for drug delivery systems~\citep{Herrmann2021}.

\section{Conclusions}\label{sec5}
We have explored the dynamical viscoelasticity of a $2$D fluid membrane featuring lipid bilayers  under oscillatory tensile loadings.
Following a previous theoretical and numerical study by \cite{Fournier2015},
we have used hydrodynamical equations of bilayer membranes.
Our numerical results show that membrane characteristics shift from elastic dominant to viscous dominant ($E^\prime < E^{\prime\prime}$) when the frequency $f$ increases.
Thus, our study provides conclusive evidence that a $2$D fluid membrane exhibits state transition depending on the oscillatory loading frequencies.
Calculated complex moduli are consistent with previous experimental measurements in both low and high frequencies~\citep{Al-Rekabi2018, Choi2011, Kim2011}.
In particular, calculated complex moduli obtained with a specific surface tension ($\sigma = 1$ mN/m) agree well with those in a previous experimental work using AFM and DPPC bilayers,
at high frequencies~\citep{Al-Rekabi2018}.
Our numerical results obtained with the width of loading profile $w = 1$ $\mu$m further show that state transition appears within the range $40$ Hz $\leq f \leq$ $400$ Hz almost independently of surface tension $\sigma$.
The transition shifts to a lower frequency range as the width of loading profile increases.
These numerical results provide fundamental knowledge to build more rigorous continuum membrane models that consider multi-scale dynamics,
and yield insight into characteristic cell dynamics at various time scales.

\section*{Credit author statement}
Naoki Takeishi, Masaya Santo, Naoto Yokoyama: data curation, figure construction, validation,
Naoto Yokoyama, Shigeo Wada: advisor, conceptualization,
Naoto Yokoyama: Methodology,
Naoki Takeishi, Naoto Yokoyama: reviewing and editing.

\section*{Declaration of competing interest}
The authors declare that they have no known competing financial interests or personal relationships that could have appeared to influence the work reported in this paper.

\section*{Acknowledgements}
This research was supported by JSPS KAKENHI Grant Number JP20H04504.
N.T. is grateful for the financial support of UCL-Osaka Partner Funds.
The presented study was partially funded by Daicel Corporation.
Last but not least, N.T. thanks Mr. Yu Taniguchi, Dr. Hiroshi Yamashita, and also Prof. Masako Sugihara-Seki for helpful discussions.

\appendix
\section{Derivation of the dynamical equation}\label{secA1}

Let us take a total differential form of the solvent velocity $V_i^\pm = V_i^\pm (\v{r}, h(\v{r}))$, ($i = x$ and $y$),
\begin{align}
	  dV_i^\pm &= \frac{\partial V_i^\pm}{\partial r_i} dr_i + \frac{\partial V_i^\pm}{\partial h} dh, \nonumber\\
	  \frac{\partial V_i^\pm}{\partial h}\frac{\partial h}{\partial r_i} &= 0, \nonumber\\
	  \frac{\partial V_i^\pm}{\partial z} \partial_i h &= 0 \quad (\because~h = z, \ \partial/\partial r_i = \partial_i).
	  \label{dV}
\end{align}
Thus, we obtain Eq.~\eqref{partial_v} from the no-slip boundary condition, Eq.~\eqref{no-slip_xy}:
\begin{align}
	v_i^{\pm} &= V_i^{\pm}|_{z = h}, \quad (\because~\eqref{no-slip_xy}) \nonumber\\
	\to \partial_i v_i^\pm
	&= \partial_i V_i^\pm|_{z = h} + \left. \frac{\partial V_i^\pm}{\partial z} \right|_{z = h} \partial_i h, \nonumber\\
	&= \partial_i V_i^\pm|_{z = h} \quad (\because~\eqref{dV}), \nonumber\\
	&= - \left. \frac{\partial V_z^\pm}{\partial z} \right|_{z = h} \quad (\because~\eqref{mass_flu}).
	\label{partial_v_ver2}
\end{align}
Finally we have Eq.~\eqref{strain_zz} as:
\begin{align}
	  \partial_t \rho^\pm
	  &= - \partial_i v_i^\pm, \quad (\because~\eqref{mass_mem}) \nonumber\\
	  &= \left. \frac{\partial V_z^\pm}{\partial z} \right|_{z = h}, \quad (\because~\eqref{partial_v_ver2}), \nonumber\\
	  \to \rho^\pm &= \int \left. \frac{\partial V_z^\pm}{\partial z} \right|_{z = h} dt.
\end{align}

Taking into account the constraints described in~\eqref{mass_flu}--\eqref{no-slip_z},
and introducing the Lagrange multiplier fields $P^\pm (\v{R})$, $\zeta^\pm (\v{r})$, $\mu_i^\pm (\v{r})$, and $\gamma^\pm (\v{r})$,
we take an extremum for the functional $\m{R}^\ast$
\begin{align}
	\m{R}^\ast
	= &\sum_{\epsilon = \pm} \int_{B^\epsilon} d\v{R} \left[ \eta D_{\alpha\beta}^\epsilon D_{\alpha\beta}^\epsilon - P^\epsilon \partial_\alpha V_\alpha^\epsilon \right] \nonumber \\
	+ &\sum_{\epsilon = \pm} \int d \v{r}
	\left[
	  \eta_2 d_{ij}^\epsilon d_{ij}^\epsilon + \frac{\lambda_2}{2} d_{ii}^\epsilon d_{jj}^\epsilon
	  + \zeta^\epsilon \left( \dot{\rho}^\epsilon + \partial_i v_i^\epsilon \right) \right. \nonumber \\
	  &\hspace{0.2cm} + \mu_i^\epsilon \left( v_i^\epsilon - V_i^\epsilon \right) + \gamma^\epsilon \left( \dot{h} - V_z^\epsilon \right) \nonumber \\
	  &\hspace{0.2cm} \left. + k \left( \rho^\epsilon + \epsilon e \nabla^2 h \right) \dot{\rho}^\epsilon
         + \epsilon k e \nabla^2 \left( \rho^\epsilon + \epsilon e \nabla^2 h\right) \dot{h}
       \right] \nonumber \\
	+ &\int d \v{r}
	  \left[
	  \frac{b}{2} \left( \v{v}^+ - \v{v}^- \right)^2 \right.\nonumber \\
	  &\hspace{0.2cm} \left. + \left( \kappa \nabla^4 h - \sigma \nabla^2 h + p_0 \exp{\left( -12|\v{r}|^2/w^2 \right)}
	  \right) \dot{h} \right].
         \label{R_ast}
\end{align}
Taking an extremum for $\m{R}^\ast$ with respect to the fields $V_\alpha^\pm$, $\dot{\rho}^\pm$, $\dot{h}$, $v_i^\pm$, $P^\pm$, $\zeta^\pm$, $\mu_i^\pm$, and $\gamma^\pm$ yields
\begin{align}
	&\partial \m{R}^\ast/\partial V_i^\pm (\v{r}, 0) = 0 \nonumber\\
	&\to \mp \eta \left( \partial_z V_i^\pm + \partial_i V_z^\pm \right) - \mu_i^\pm = 0, \\
	&\partial \m{R}^\ast/\partial V_z^\pm (\v{r}, 0) = 0 \nonumber\\
	&\to \mp 2 \eta \partial_z V_i^\pm \pm P^\pm - \gamma^\pm = 0, \\
	&\partial \m{R}^\ast/\partial \dot{\rho}^\pm = 0 \nonumber\\
	&\to \zeta^\pm + k \left( \rho^\pm \pm e\nabla^2 h \right) = 0, \\
	&\partial \m{R}^\ast/\partial \dot{h}^\pm = 0 \nonumber\\
	&\to \gamma^\pm + \tilde{\kappa} \nabla^4 h - \sigma \nabla^2 h + k e \nabla^2 \left( \rho^+ - \rho^-\right) \nonumber\\
	&\quad + p_0 \exp{\left( -12|\v{r}|^2/w^2 \right)} = 0, \label{elastic_force}\\
	&\partial \m{R}^\ast/\partial v_i^\pm = 0 \nonumber\\
	&\to \eta_2 \partial_j \partial_j v_i^\pm - \left( \eta_2 + \lambda_2 \right) \partial_i \partial_j v_j^\pm	- \partial_i \eta^\pm + \mu_i^\pm \nonumber \\ 
	&\quad + b \left( v_i^\pm - v_i^\mp \right) = 0.
\end{align}
For further description of the derivation of the membrane hydrodynamic equations in Eqs.~\eqref{governing_e1} and \eqref{governing_e2}, see Ref.~\citep{Fournier2015}.

\section{Linear Maxwell materials}\label{secA2}
The constitutive equations for the Maxwell materials,
which are represented by a linear combination of the two types of material responses, specifically a dashpot (viscous fluid) and a spring (elastic solid),
can be written as
\begin{equation}
	\Sigma (t) + \frac{\eta_d}{G} \dot{\Sigma} (t) = \eta_d \dot{\varepsilon} (t),
	\label{Maxwell}
\end{equation}
where $\Sigma (t)$ and $\varepsilon (t)$ are the total stress and strain at time $t$,
$G$ is the elastic constant of the material,
and $\eta_d$ is the fluid viscosity.
The time derivations of periodic strain $\varepsilon (t) = \varepsilon_0 \exp{ \left( i \omega t \right)}$ and its response $\Sigma (t) = \Sigma_0 \exp{\left( i \omega t + \delta \right)}$ are written as
\begin{align}
	\dot{\varepsilon} (t) &= i \omega \varepsilon_0 \exp{ \left( i \omega t \right)} = i \omega \varepsilon (t), \\
	\dot{\Sigma} (t) &= i \omega \Sigma_0 \exp{\left( i \omega t + \delta \right)} = i \omega \Sigma (t),
\end{align}
Substituting these equations into Eq.~\eqref{Maxwell},
we have
\begin{align}
	&\Sigma (t) + i \omega \frac{\eta_d}{G} \Sigma (t) = i \omega \eta_d \varepsilon (t), \nonumber \\
	\to &\Sigma (t) = \left\{ \frac{(\omega \m{T})^2 G}{1 + (\omega \m{T})^2} + i \frac{\omega \m{T} G}{1 + (\omega \m{T})^2} \right\} \varepsilon(t), 
	\label{form_1} \\
	&\hspace{0.7cm} = \frac{(\omega \m{T})^2 G}{1 + (\omega \m{T})^2} \varepsilon (t) + \frac{\eta_d}{1 + (\omega \m{T})^2} \dot{\varepsilon} (t),
	\label{form_2}
\end{align}
where we have introduced the Maxwell relaxation time $\m{T} = \eta_d/G$~\citep{Mewis2012}.
Equations~\eqref{form_1} and \eqref{form_2} demonstrate that the Maxwell model exhibits a stress response both in and out of phase with the applied deformation.
Comparing the expressions in Eq.~\eqref{linear_response} and Eq.~\eqref{form_1},
we conclude that complex modulus $G^\ast (= G^\prime (\omega) + i G^{\prime\prime} (\omega))$ is defined as:
\begin{align}
	&G^\prime (\omega) = \frac{(\omega \m{T})^2}{1 + (\omega \m{T})^2} G, \\
	&G^{\prime\prime} (\omega) = \frac{\omega \m{T}}{1 + (\omega \m{T})^2} G.
\end{align}
We also conclude that the stress response can be interpreted in terms of a frequency-dependent elastic modulus $\tilde{G}$ and viscosity $\tilde{\eta}$ as
\begin{align}
	&\tilde{G} (\omega) = G^\prime (\omega) = \frac{(\omega \m{T})^2 }{1 + (\omega \m{T})^2} G, \\
	&\tilde{\eta} (\omega) = \frac{1}{1 + (\omega \m{T})^2} \eta_d
\end{align}
At short times ($\omega \m{T} \gg$ 1),
the Maxwell model behaves like a solid with the elastic modulus $\tilde{G} (\omega) \approx G$,
while at long times ($\omega \m{T} \ll$ 1) it behaves as a viscous fluid with viscosity $\tilde{\eta} (\omega) \approx \eta_d$.
The crossover between the two states occurs when the time scale of deformation is similar to the time scale of relaxation, $\omega^{-1} \sim \m{T}$.

\printcredits

\bibliographystyle{cas-model2-names}

\bibliography{cas-refs}

\end{document}